\documentstyle[12pt,aasms4]{article}

\lefthead{Bekki, Shioya, \& Couch}
\righthead{Formation of E+A galaxies}

\begin{document}
\title{Formation and evolution of E+A galaxies in dusty starburst galaxies}

\author{Kenji Bekki} 
\affil{
School of Physics, University of New South Wales, Sydney 2052, Australia}

\author{Yasuhiro Shioya} 
\affil{
Astronomical Institute, 
Tohoku University, Sendai, 980-8578, Japan}

\and

\author{Warrick J. Couch}
\affil{
School of Physics, University of New South Wales, Sydney 2052, Australia}

\begin{abstract}
The formation and evolution of the ``E+A'' (also named ``k+a'' and 
``a+k'' types by Dressler et al. 1999) galaxies found in significant
numbers in the cores of intermediate redshift clusters has been 
extensively discussed by many authors. In this paper, we model the
spectral, dynamical and morphological evolution of a prime candidate
for producing this spectral signature: a dusty starburst associated with
a major galaxy merger. We show that as this system evolves dynamically,
its spectral type changes from and ``e(a)'' type (exhibiting strong
H$\delta$ absorption and modest [OII] emission -- the identifying 
features of local dusty starburst galaxies) to a k+a type and then
finally to a passive ``k'' type. This result shows that galaxies with
an e(a) spectral type can be precursors to the k+a systems and 
that dynamical evolution greatly controls the spectral evolution 
in these merger cases. Our simulations also show that a merger with very
high infrared luminosity ($L_{\rm IR}$ $>$ $10^{11}$ $L_{\odot}$) is more
likely to show an e(a) spectrum, which implies that spectral types can be
correlated with infrared fluxes in dusty starburst galaxies. Based on
these results, we discuss the origin of the evolution of k+a/a+k
galaxies in distant clusters and the role merging is likely to have.
\end{abstract}

\keywords{galaxies: clusters: general ---  galaxies: ISM --- 
galaxies: elliptical and lenticular, cD -- galaxies: formation --
galaxies:
interactions
}

\section{Introduction}

The origin of the so-called ``E+A'' galaxies with no detectable emission
and strong Balmer absorption lines, is generally considered to be one of
the most longstanding and remarkable problems of galaxy evolution in
distant ($z>0.2$) clusters. Dressler \& Gunn (1983) proposed that this
peculiar spectroscopic property was due to the presence of a substantial
population of A-type stars, having been formed as part of a recent
starburst which was abruptly truncated. Since Couch \& Sharples (1987) 
suggested a possible evolutionary link between these E+A galaxies
and the general population of blue galaxies found in distant clusters
(Butcher \& Oemler 1978), several attempts have been made to determine
what physical processes are closely associated with the abrupt truncation 
of star formation that is inferred from their spectral characteristics
(e.g., Abraham et al. 1996; Barger et al. 1996; Balogh et al. 1997; 
Couch et al. 1998; Poggianti et al. 1999). Although the formation of E+A
galaxies has been often discussed in terms of physical processes specific
to the cluster environment, Zabludoff et al. (1996) found that a large 
fraction of nearby E+A galaxies lie in the field rather than in clusters
and therefore suggested that cluster environmental effects such as 
interaction with the cluster gravitational potential or intracluster
medium are not responsible for E+A formation. 

Recent detailed morphological and spectroscopic studies by the $Hubble$ 
$Space$ $Telescope$ ($HST$) and large ground--based telescopes have shed
new light on the origin of E+A galaxies. For example, a significant fraction 
of the distant cluster galaxies with E+A spectra -- renamed by Dressler et
al. (1999) as ``a+k'' or ``k+a'' types -- are observed to be disk systems,
which implies that abrupt (or even gradual) truncation of star formation 
(after the starburst) has occurred without a dramatic transformation in
their morphology (Couch et al. 1998; Dressler et al. 1999). Furthermore, 
Poggianti et al. (1999) argued on the basis of the observed fractions
and luminosity functions of the different spectral types that the 
progenitors of a+k/k+a galaxies were the ``e(a)'' class objects -- those
with modest [OII] emission and strong Balmer line absorption 
[EW(H$\delta)>4$\AA], the same two features that characterize the spectra
of nearby dusty starburst galaxies (Liu \& Kennicutt 1995). 
These observational results strongly suggest that it is very important for
theoretical studies to address the following three problems: (1)\,What
are the physical processes responsible for the formation of E+A galaxies?
(2)\,What are the evolutionary links between the different spectral types, 
in particular the e(a), e(b), e(c), a+k, k+a, and k class objects
identified by Dressler et al. (1999)? (3)\,How is the observed
morphological evolution (e.g., the smaller fraction of S0 populations in
higher redshift clusters; Dressler et al. 1997) related to spectral
evolution of galaxies in clusters? Clearly both the morphological and
spectral properties of distant cluster galaxies have to be investigated
{\it jointly} in a fully self-consistent manner if these questions are to
be answered. However, previous theoretical studies have investigated
quite separately the morphological transformation processes (using
numerical methods; Byrd \& Valtonen 1990; Moore et al. 1994) and the 
possible spectral evolutionary links (using simple one-zone models; 
Barbaro \& Poggianti 1997; Poggianti et al. 1999; Shioya \& Bekki 2000).
Thus, it is still highly uncertain when and how E+A galaxies form and
evolve both spectrally and morphologically.

The purpose of this Letter is to investigate the morphological, photometric, 
and spectroscopic properties of galaxies in an explicitly self-consistent
manner and thereby provide some plausible and realistic answers to the
above three problems. We adopt an observationally suggested scenario that
$some$ E+A galaxies are formed by strong interactions and merging (e.g.,
Zabludoff et al. 1996), and demonstrate when and how a dusty starburst
triggered by a major galaxy merger develops an E+A spectrum in the course
of its dynamical evolution. Here we use our newly developed model (Bekki
\& Shioya 2000, hereafter referred to as BS) by which we can
self-consistently investigate the time evolution of spectral types (e.g.,
k+a, a+k, e(a)... etc), infrared fluxes (e.g., $L_{\rm IR}$ for 8$-$1000\,
$\mu$m), and dynamical properties (e.g., radial density profile) in a
galaxy with dusty starbursts. Based on
our new `spectrodynamical' simulations  (BS), we particularly demonstrate that a
merger exhibits a k+a/a+k spectrum following its strong starburst phase 
and, furthermore, that the dynamical evolution of a galaxy can greatly
control its spectral evolution. 

\section{The model}

Since our numerical methods and techniques for modeling chemodynamical
and photometric evolution of dusty starbursts associated with major galaxy
mergers have already described in detail by BS, we give only a brief
review here. We construct models of galaxy mergers between gas-rich 
disks with equal mass by using the model of Fall-Efstathiou (1980).
The total mass and the size of a progenitor exponential disk are 
$M_{\rm d}$ and $R_{\rm d}$, respectively. From now on, all the mass and 
length are measured in units of $M_{\rm d}$ and  $R_{\rm d}$,
respectively, unless otherwise specified. Velocity and time are 
measured in units of $v$ = $ (GM_{\rm d}/R_{\rm d})^{1/2}$ and
$t_{\rm dyn}$ = $(R_{\rm d}^{3}/GM_{\rm d})^{1/2}$, respectively,
where $G$ is the gravitational constant and assumed to be 1.0
in the present study. If we adopt $M_{\rm d}$ = 6.0 $\times$ $10^{10}$ $
\rm M_{\odot}$ and $R_{\rm d}$ = 17.5 kpc as a fiducial value, then $v$ =
1.21 $\times$ $10^{2}$ km/s  and  $t_{\rm dyn}$ = 1.41 $\times$ $10^{8}$
yr, respectively. The disk-to-dark halo mass ratio is set equal to
4.0. The total mass of gas is 3.0 $\times$ $10^{10}$ $ \rm M_{\odot}$.
The orbital configuration of the present major merger between
the above two disks is exactly the same as that in BS.
To calculate spectral energy distributions (SEDs) 
of the merger and luminosities of gaseous emission lines, e.g., 
[OII] and H$\delta$, we use the spectral library GISSEL96 (Bruzual \& Charlot (1993)
and the formula of 
$L({\rm H}\beta)({\rm erg \; s^{-1}})=4.76 \times 10^{-13}
N_{\rm Ly}({\rm s}^{-1})$, where $N_{\rm Ly}$ is 
ionizing photon production rate, and
the table of relative luminosity to H$\beta$ luminosity 
in PEGASE (Fioc \& Rocca-Volmerange 1997). 
We investigated both a model without dust (referred to as ``DF'' for dust
free) and a model with dust (``DS'' or dusty starburst).

\placefigure{fig-1}
\placefigure{fig-2}

\section{Results}

Figure 1 describes the time evolution of EW([OII]) and EW(H$\delta$) of
the merger-induced starburst over the period $0.6\le T\le 2.8$\,Gyr for
the DF and the DS models. At the point where the star formation rate
rapidly increases ($T\sim 1.3$\,Gyr), EW([OII]) is also seen to rapidly 
increase ($100\le$ EW([OII]) $\le 370$\AA) in the DF model. In contrast, 
the DS model does not show such a dramatic increase during the starburst
phase owing to the very heavy dust extinction. The DS model shows 
EW([OII]) does not exceed 40\AA\ and the difference in EW([OII]) between the 
two models is rather large, ranging from a factor of $\sim$ 3 at 
$T = 1.1$\,Gyr to a factor of 35 at $T = 1.3$\,Gyr.
After the starburst ($T>1.3$\,Gyr), EW([OII]) rapidly decreases
with time for the two models. 

Owing to the heavy dust extinction, EW(H$\delta$) does not show a large
negative value during the starburst and is reasonably well within the
observed values ($<10$\,\AA; Liu \& Kennicutt 1995) of dusty starburst
galaxies for the DS model. The most important point in EW(H$\delta$) evolution 
is that only the DS model shows an absorption line signature (i.e.,
positive values) during the starburst and poststarburst epochs: {\it Dust
effects are critically important for the EW(H$\delta$) evolution in dusty
galaxies!}

Figure 2 demonstrates that as a natural result of the above EW([OII]) and
EW(H$\delta$) evolution, the spectral type of the DS model evolves from
e(b) ($T = 0.6$\,Gyr), to e(a) (1.3), to k+a (1.7), and finally to 
k (2.8). This result confirms that a dusty starburst associated with
a major merger can become a k+a galaxy $\sim$ 0.4 Gyr after its burst
of star formation. One of the remarkable differences in the spectral
evolution on the EW([OII])-EW(H$\delta$) plane between the two models is
that only the DS model shows the e(a) spectral signature to persist for 
any length of time ($\sim 0.3$\,Gyr) during the period of active star
formation.
This result suggests that formation of e(a) galaxies
is closely associated with dust extinction, thus confirming
the earlier suggestion by Poggianti et al. (1999) and Shioya \& Bekki
(2000). Furthermore, as is shown in Figure 3, our modeling approach allows
us to clearly observe the different morphological phases of the merger
and to track the changes in spectral type in tandem through each.  
For the e(b) spectral phase, two tidally interacting disks are clearly
observed whereas for the e(a) spectral phase, only a morphologically
peculiar galaxy with clear signs of merging (a tidal dwarf and a very diffuse 
plume-like structure) can be seen. For the k+a/k phases, the merger
becomes almost dynamically relaxed and thus does not show any clear
relic of the past merging event(s). These results imply that
the observed fraction of mergers among a+k/k+a (i.e., E+A) galaxies
(e.g., Zabludoff et al. 1996) can actually be underestimated. 

Figure 4 clearly shows an interesting physical correlation between the
evolution in the infrared luminosity ($L_{\rm IR}$) and that seen in H$\alpha$ 
emission. To be more specific, as the $L_{\rm IR}$ becomes higher,
H$\alpha$ emission becomes also stronger during the evolution
in the DS model.
The present  model furthermore  shows
the larger difference in H$\alpha$ luminosity between the two models (DF and DS)
when the $L_{\rm IR}$ becomes higher
(e.g., a factor of 100  difference for $T$ = 1.3 Gyr). 
Considering the results given in Figure 2, these  results suggest that
(1)\,a dusty starburst merger with an e(a) spectrum can exhibit very high
$L_{\rm IR}$ ($>$ $10^{11}$ $L_{\odot}$), but (2)\,this might not 
necessarily be accompanied by a high luminosity in H$\alpha$. The first
of these suggestions is consistent with recent observations of infrared
luminous e(a) galaxies by Poggianti \& Wu (2000). The second means that
even H$\alpha$ observations can considerably underestimate (by a factor
of $\sim 100$) the real star formation rate of a dusty starburst galaxy,
and therefore infrared observations are much better for inferring
the true star formation rate. Thus our spectrodynamical model has not only
succeeded in providing an evolutionary link between galaxies with
different spectral types (e.g., e(a) and k+a) but has also suggested
physical correlations between spectral types, morphological properties,
and infrared fluxes.

\placefigure{fig-3}
\placefigure{fig-4}

\section{Discussion and conclusion}

 Our numerical results prompt the following
question: can major galaxy merging account for the observation that the
fraction of k+a/a+k galaxies in distant clusters (10-20\% at 
$z\sim 0.4$; e.g., Dressler et al. 1999) is an order of magnitude
higher than the fraction observed in low redshift clusters? 
From their own observations of the cluster MS1054-03 at $z=0.83$ together
with data taken from studies at lower redshift, van Dokkum et al. (1999) 
showed that the fraction of interacting/merging galaxies evolved very rapidly
with redshift as ${(1+ \rm z)}^{6\pm2}$ over the interval $0\le z\le 0.83$. 
By using this observed relation and assuming (i)\,that it refers to
$major$ mergers, and (ii)\,that all these mergers evolve spectrally 
to become k+a/a+k galaxies, the expected fraction of k+a/a+k galaxies
($f_{\rm k}$) is found to be 7.5 \% ($3.8\%\le f_{\rm k}\le 14.8\%$)
at $z=0.4$. The expected value is similar to the observed one 
(10-20\%), which is consistent with major merging being the physical
mechanism which drives the rapid evolution of the k+a/a+k population   
(We here stress that since the merger rate at high redshift is estimated
only for the $single$  $z=0.83$ cluster, the expected merger rate 
at intermediate redshift is
not so reliable). 
However, this conclusion is strongly countered by the observation that a
significant fraction of the k+a/a+k types are {\it disk} systems
(e.g., See Figure  10 in Dressler et al. 1999) -- an unlikely candidate
for the type of system likely to be produced in a major merger! Indeed, 
only a {\it minor} merger would seem capable of keeping the disk component
intact. Thus, considering the observed redshift evolution of merger rates,
that of a+k/k+a populations, and the dependence of the morphological
transformation processes on the mass ratio of the two merging disks, it 
would seem that major merging can explain the origin of only a small
fraction of the k+a/a+k population.   

We have found that dusty starburst galaxies with e(a) spectra are 
plausible precursors to a+k/k+a galaxies. This raises a further question:
what fraction of the k+a/a+k galaxies observed in distant clusters were
previously starburst galaxies with e(a) spectra? Although a comparison of
the fraction of e(a) galaxies with that of k+a/a+k types gives a rough
estimate of the fraction of k+a/a+k galaxies which have evolved from
e(a) galaxies (Couch \& Sharples 1987; Barger et al 1996; Poggianti et
al. 1999), the dusty nature of starburst galaxies makes this estimation
considerably inaccurate. For example, based on the detection of some k+a
galaxies by 1.4 GHz VLA radio observations, Smail et al. (1999) suggested
that even k+a galaxies classified by EW([OII]) and EW(H$\delta$) can be
dusty starburst galaxies. Hence more accurate observational
estimation of star formation rate is necessary for answering the above
question. 

Couch et al. (2000) have investigated the incidence of H$\alpha$ emission
-- considered to be much less affected by dust than [OII] emission --
in members of the cluster AC114 at $z = 0.32$ and found that the star
formation rate in the $\sim 5\%$ of galaxies so found to be strongly
and uniformly suppressed (less than 4 $M_{\odot}$ ${\rm yr}^{-1}$)
right throughout the cluster ($r<2.25\,h_{50}^{-1}$\,Mpc).
Balogh \& Morris (2000) also investigated  H$\alpha$ emission from
galaxies in Abell 2390 at z = 0.23 and also found the star formation
rate inferred from H$\alpha$ to be rather low.
These observational studies strongly suggest that, unless the dust
extinction is  sufficient to totally obscure the H$\alpha$ emission (which
would require 4 magnitudes of extinction; Couch et al. 2000), the fraction
of ongoing dusty starbursts must be very small, as would also be the
fraction of galaxies which evolve from e(a) into k+a/a+k's. 
However, considering the observational evidence that the star formation 
rate inferred from H$\alpha$ emission is a factor of 5--20 times smaller 
than that from far-infrared fluxes for strongly star-bursting dusty  galaxies 
(e.g., Poggianti \& Wu 2000), more systematic infrared and radio
observations will be important for estimating the true fraction of ongoing
starburst galaxies in distant clusters. It will only be the combination
of systematic H$\alpha$, infrared, and radio observations that will
unambiguously measure the star formation rate in dusty star-forming
galaxies and this answer the question as to the origin of k+a/a+k
galaxies.

\acknowledgments

Y.S. thanks the Japan Society for Promotion of Science (JSPS) 
Research Fellowships for Young Scientists.

\clearpage


\figcaption{
$Top$:  Time evolution of the core separation of two merging disks in units of kpc.
Star formation rate becomes maximum ($\sim$ 3 $\times$ $10^2$ $M_{\odot}$ ${\rm yr}^{-1}$)
when the two cores (disks)  finally merge to form a single core of an elliptical galaxy
($T$ = 1.3 Gyr).
Star formation is  0 $M_{\odot}$ ${\rm yr}^{-1}$
at $T$ $>$ 1.4 Gyr when the two disks become an elliptical galaxy. 
$Middle$: Evolution of the equivalent width of [OII] emission
line [EW([OII])]. The DF model (the dust free model, without dust extinction)
and the DS one (the dusty starburst model, with dust extinction) 
are shown as the dotted and solid lines, respectively.
Here the time $T$ represents time that has elapsed since
the two disks begin to merge.
$Bottom$: Evolution of the equivalent width of H$\delta$ [EW(H$\delta$)]
for the DF model (dotted) and the DS one (solid). 
\label{fig-1}}

\figcaption{
$Top$: Evolution of galaxies on the EW([OII])-EW(H$\delta$) plane
for the ranges of 0 $\le$ EW([OII]) $\le$ 500 $ \rm \AA$
and  $-40$ $\le$ EW(H$\delta$)  $\le$ 10 $\rm \AA$. The DF
and DS models are shown as dotted and solid lines, respectively.
The results at $T$ = 0.56, 1.12, 1.24, 1.30, 1.34, 1.39, 1.69,
2.26, and 2.82 Gyr are given and connected with each other along the time sequence. 
$Bottom$: The same as the top panel 
but for the ranges of 0 $\le$ EW([OII]) $\le$ 200 $ \rm \AA$
and $-5$ $\le$ EW(H$\delta$)  $\le$ 10 \AA. 
The criteria of spectral classification by Dressler et al (1999)
are superimposed.
For the two models,
the initial ($T$ = 0.56 Gyr) and final (2.82) spectral types are   estimated to be 
e(b) and k, respectively. 
Note that only the DS model shows an e(a) spectrum for a fairly long time scale
($\sim$ 0.3 Gyr). 
Note also that the DS model evolves from e(b), to e(a), to k+a, and to k.
This figure moreover demonstrates  that if the  effects of dust  are  negligible,
the merger can show an a+k spectrum after undergoing a dusty starburst (Compare  DF model with
DS one).
\label{fig-2}}

\figcaption{
Morphological properties and
spectral types for  the dusty starburst merger projected onto $x$-$y$ plane
at $T$ = 0.56 Gyr (upper left), 1.30 (upper right), 1.69 (lower left),
and 2.82 (lower right). Each frame measures 68.3 kpc on a side.
Spectral types (e.g., e(a)) and time ($T$)
are indicated  in the upper left-hand  corner for each frame. 
\label{fig-3}}

\figcaption{
Time evolution of infrared luminosity ($L_{\rm IR}$) 
for the DS model (short-dashed), H$\alpha$ luminosity for the DF one (dotted),
and that for the DS one (solid) in units of $W$.
Here $L_{\rm IR}$ represents the total luminosity from dust reemission
of two merging disk galaxies.
\label{fig-4}}

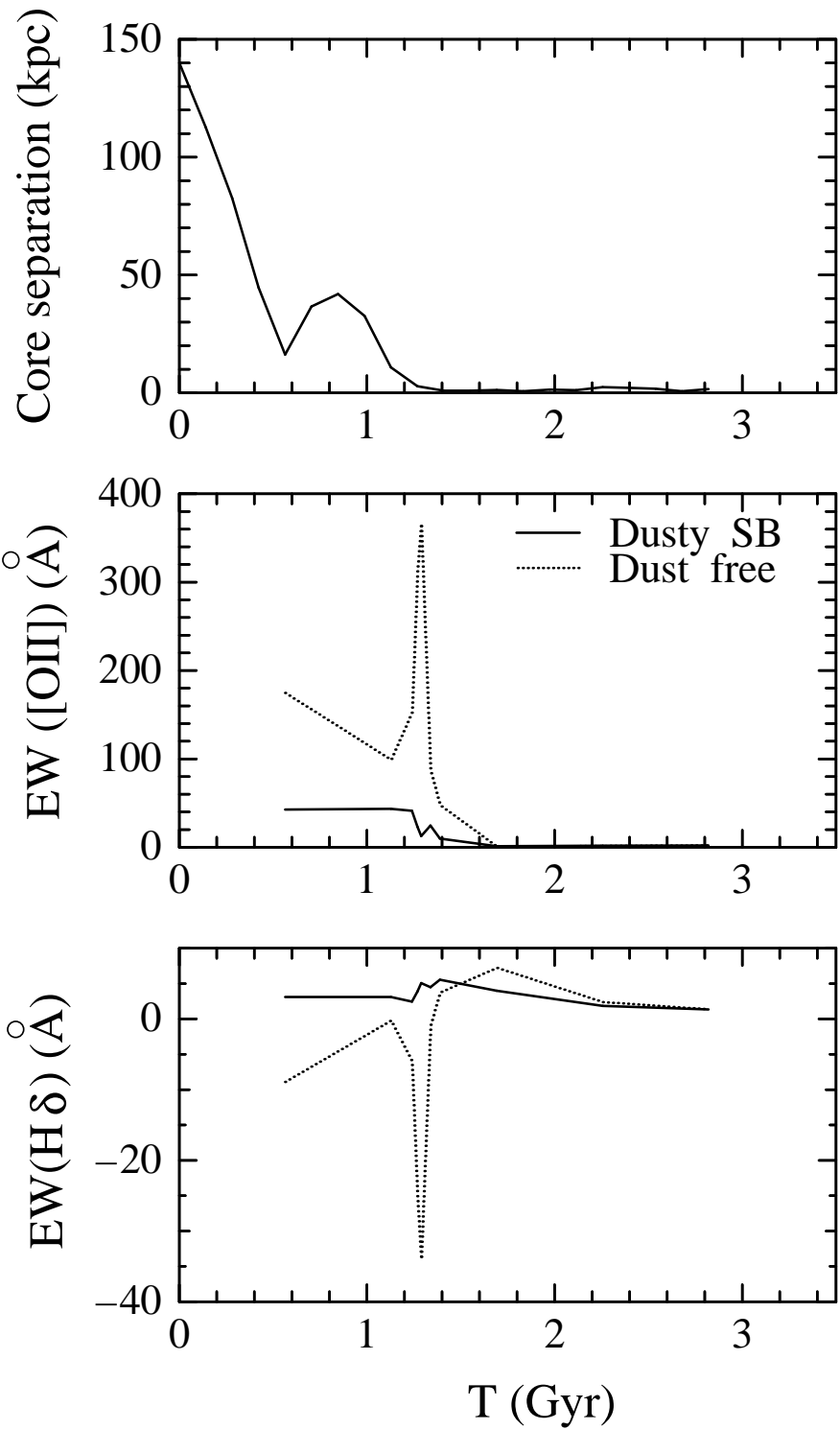
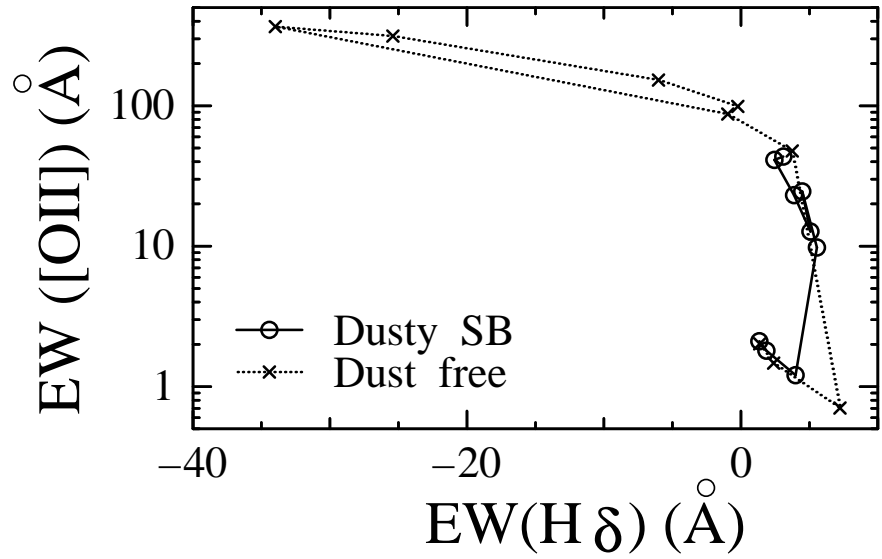
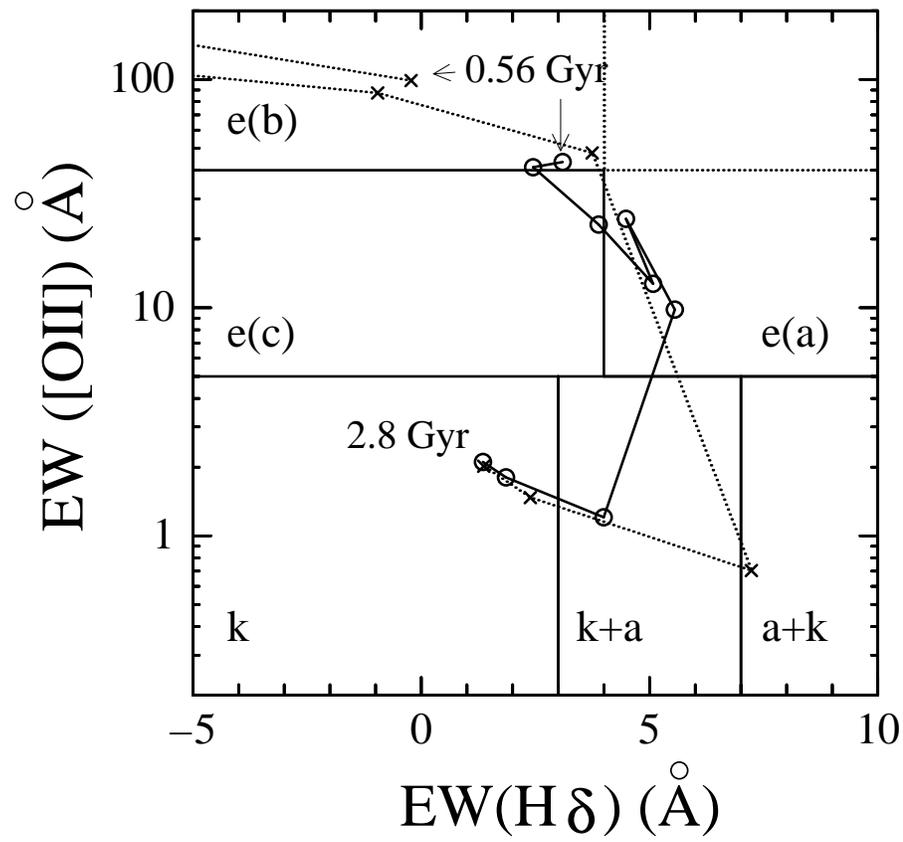
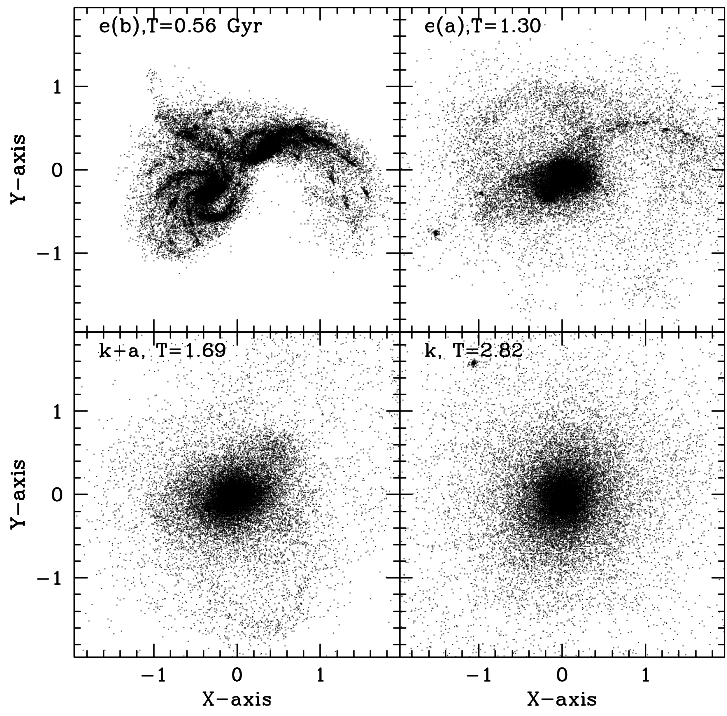
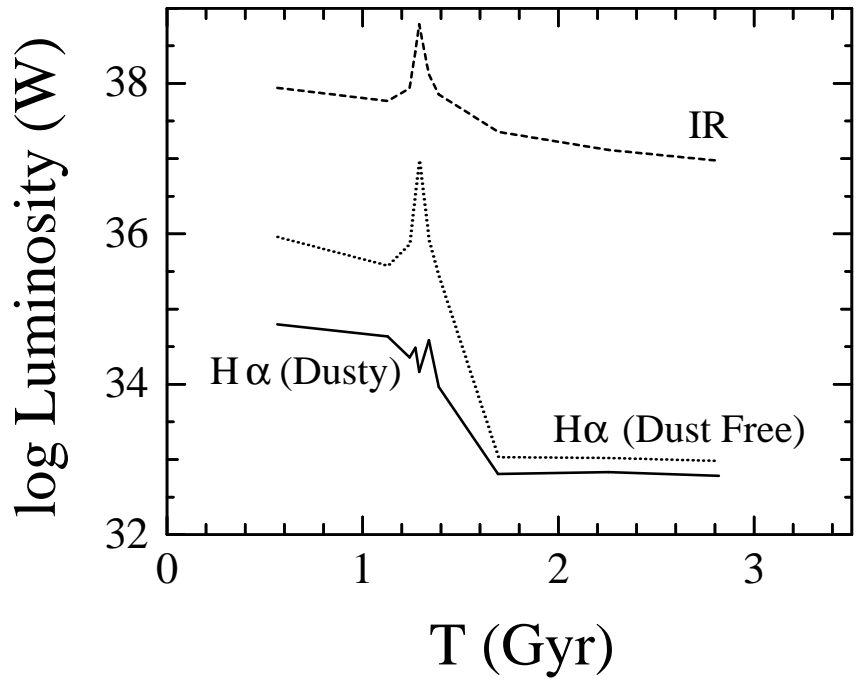
\end{document}